# Knowledge Graph Development for App Store Data Modeling

*Mariia Rizun*  mariia.rizun@ue.katowice.pl
University of Economics in Katowice
Katowice, Poland

*Artur Strzelecki*  artur.strzelecki@ue.katowice.pl
University of Economics in Katowice
Katowice, Poland

**Abstract**

Usage of mobile applications has become a part of our lives today, since every day we use our smartphones for communication, entertainment, business and education. High demand on apps has led to significant growth of supply, yet large offer has caused complications in users' search of the one suitable application. The authors have made an attempt to solve the problem of facilitating the search in app stores. With the help of a website crawling software a sample of data was retrieved from one of the well-known mobile app stores and divided into 11 groups by types. These groups of data were used to construct a Knowledge Schema – a graphic model of interconnections of data that characterize any mobile app in the selected store. Schema creation is the first step in the process of developing a Knowledge Graph that will perform applications clustering to facilitate users' search in app stores.

## 1. Introduction

In the present time knowledge management has been gaining its importance in most organizations, since every day they are obtaining a lot of information, processing large massive of data and generating more and more new knowledge. In such a fast pace it becomes more difficult to control and organize all the data flows in a proper way, make the information available for all workers and enable any piece of information to be searched quickly and easily [28], [11], [14].

However, not only brick-and-mortar organizations or institutions face the need of having the knowledge structured. In World Wide Web, the world's largest information space, around 2,5 quintillion bytes of data are created every day (according to Forbes[1]); more than 5 billion searches are processed every day (with 77% being conducted on Google). It is necessary to add that more than half of all the web searches are done from smartphones or tablets. This fact is not surprising since every reader would agree that there is almost no day in our today's life that we spend without using at least one of these gadgets. Moreover, most of our interactions with smartphones are based on using mobile applications (for calling, texting, searching for information, taking notes, playing games, etc.).

Today, a user is offered a large variety of mobile applications, provided by various repositories [25], [19]: the well-known Google Play store (previously Android Market), Apple App Store, Amazon Appstore for Android, BlackBerry World store, as well as some others (e.g. MalGenome, Drebin, AndroZoo). The perk of having such a wide choice of applications is quite obvious – a user can find an application that 100% fits his/her interests and tastes, as well as his/her smartphone software requirements. Yet being offered a large variety of applications (of similar categories, of same price, with similar functions, etc.) a user gets confused and sometimes has to spend a lot of time on selecting the one particular app [23]. That is why developers of the above-mentioned repositories are continuously working on improvement of the search mechanisms in order to facilitate the process of choosing applications at their platforms and, as a result, stay valuable for users and competitive in the market. At the same time, a lot of researchers are taking their own attempts to solve the problem of improving search engines in mobile app stores.

---

[1] https://www.forbes.com

This paper describes the first step of the authors research on mobile applications clustering. The idea is based on development of a Knowledge Graph that will structure data form applications. The use case for Knowledge Graph development is Google Play store (former Android Market). The data sample has been retrieved from the store with a website crawling software. The contribution of this study to general research is the combination of Knowledge Graph development methodology with the algorithms of app store data mining.

The paper is organized as follows. Section 2 contains review of the relevant literature on app stores mining and development of Knowledge Graphs. It also sets the Research Questions to be answered in the paper. Section 3 presents the methodology of retrieving data from Google Play store and of constructing a Knowledge Schema based on these data. Section 4 contains description of the obtained data and of the knowledge Schema, which models the concept of a mobile application. In Section 5 the authors highlight the contribution of the research, draw final conclusions and present suggestions as for their future research on this topic.

## 2. Literature Review

The authors divided the literature review into three sections, that will follow successively. First, research papers on mining data from mobile application stores are analyzed, since such data mining is further applied by authors in the practical part. Second section covers development of a Knowledge Graph and construction of the Knowledge Schema as its basis. This part of analysis is relevant because the major objective of this paper is developing a Knowledge Graph for mobile applications: in the third, and last, section, the authors explore research works to reveal whether much has been done in implementing Knowledge Graphs for analyzing and clustering mobile applications.

Research papers dedicated to the issues of app store mining can be divided into two main groups based on the mining strategies they apply. The first approach is to mine all the available information about an app: its price, technical description, number of downloads, rating, etc. Suh et al. [35] analyzed and visualized the structure of smartphone application services to identify which of them are being developed and provided in detail in an App Store. Results were illustrated in three visual forms: grid, tree, and network. Harman et al. [10] used data mining to extract feature information, which then was combined with some immediately available information to analyze apps' technical, customer and business aspects. This approach was applied to 32,108 non-zero priced apps available in the Blackberry app store. Results show that there is a strong correlation between customer rating and the rank of app downloads yet, surprisingly, there is no correlation either between price and downloads, or between price and rating. Finkelstein et al. [7] used data mining to extract price and popularity information, then – natural language processing and data mining: in order to elicit each app's claimed features from the Blackberry App Store, revealing strong correlations between customer rating and popularity (rank of app downloads). They found evidence for a mild correlation between price and the number of features claimed for an app to have, and also found that higher priced features tended to be lower rated by their users. Martin et al. [21] introduced the app sampling problem and studied its effects on sets of user review data. This problem exists when only a subset of apps is studied, resulting in potential sampling bias. They found that app metrics such as price, rating, and download rank are significantly different between the sets with fully complete review data, partially complete review data, and with no review data.

The second approach to app store mining consists in conducting text mining only of reviews written by users. Reviews are a source for exploring users' feedback, requests for new features or bugs reporting. Iacob and Harrison [13] revealed that 23.3% of reviews represent feature requests, i.e. comments through which users either suggest new features for an app or express preferences for the re-design of already existing features of an app. Pagano and Maalej [29] provided exploratory study, which analyzes over one million reviews from the Apple App Store. They found that most of the feedback is provided shortly after new releases, with a quickly decreasing frequency over time. Reviews typically contain multiple topics, such as user experience, bug reports, and feature requests. Chen et al. [5] presented a novel computational framework for reviews mining, which performs comprehensive analytics from raw users' reviews by extracting informative reviews (filtering out noisy and irrelevant ones), then automatically grouping the informative reviews using topic modeling, further prioritizing the informative reviews by an effective review ranking scheme, and finally presenting the groups of most informative reviews. Lai et al. [18] analyzed 4,480 user feedbacks from a health and fitness-tracking app in the Google Play, using text mining. The result of this study shows that users of health and fitness-related apps are concerned about their physical activity records and physiological records. The records include track, distance, time, and calories burned during jogging or walking. App store reviews are used to analyze different aspects of app development and evolution. However, users' feedback does not only exist on the app store. Nayebi et al. [27] studied how Twitter can provide complementary information to support mobile app development. Genc-Nayebi and Abran [8] did a systematic literature review of opinion mining studies from mobile app store user reviews. Martin et al. [22] created a review which describes and compares the areas of research that have been explored in app store analysis so far, drawing out common aspects. They analyzed 187 papers connected with the term „app store analysis„.

When moving further to exploring the phenomenon of the Knowledge Graph (hereafter – KG or Graph), it is necessary to begin with discussing a Knowledge Schema (hereafter – KS or Schema), which [34]: forms the meta-layer for a Knowledge Graph and defines its internal structure, keeping similar classes of entities in abstract containers; contains potential relationships between classes and entities; is referred to as a reference point for integrating new data or constructing new queries.

For KS construction (and further data processing) it is possible to use Resource Description Framework Schema[2] (RDF), which provides a means to define vocabulary, structure and constraints for expressing meta data of web resources. RDF forms a simple data model and a standardized syntax for meta data. It can be considered a language for writing down factual statements [2]. In accordance with the RDF standard, information in KS (and its technical description) is represented in so-called "triples" (subject-predicate-object), where the two entities (subject and object) are related to each other by a predicate. Each triple indicates a particular fact, showing interrelations of two selected entities.

When creating triplets for any knowledge domain, it is necessary to have a formed vocabulary of entities and their relations. For this purpose, standard vocabularies are recommended, since they have already been tested in many iterations, possess similar semantics and will be advantageous in integration with other systems and data sources. If no suitable vocabulary is found, or the selected vocabulary does not contain some of the required entities, new elements can be defined, provided that they are set in semantic relation to elements from the existing standard vocabularies [15].

Many examples of purposes of constructing a Knowledge Schema can be found in research works. For instance, in the work [1], a Knowledge Schema is developed for a platform that performs mobile measurements in metrology. A Knowledge Schema that models connections of topics of Ted Talks[3] is developed in [9]. The authors of [36] construct a KS that systematizes knowledge on products, processes and production systems in the sphere of composites manufacturing. An experimental KS is developed in [24] on the dataset about Russia's culture, nature, famous people, entertainments and other features. In [15], a Schema is developed for a software company: as the basis for a Knowledge Graph that integrates knowledge from various sources to develop a service desk that will provide automated answers for customers. In [34] there is an attempt to build a KS to describe a concept of a university didactic course.

---

[2] World Wide Web Consortium [https://www.w3.org]
[3] https://www.ted.com

It can be stated that a Knowledge Graph represents interrelation of information entities, storing knowledge in accordance with the particular scheme, which is a Knowledge Schema. Yet, it is necessary to refer to the existing scientific research in order to get the full picture about Knowledge Graphs.

One of the first (and, probably, most well-known) references to the KG notion appeared back in 2012, when Google[4] introduced its Graph, presenting it as a search tool that will "help discover information quickly and easily". Google has never revealed the principles of its KG construction, yet it inspired developers to make attempts of developing new graphs (for instance, DBpedia[5], Freebase[6]) or forming vocabularies for them (e.g. Schema.org[7], presently used by more than 10 million sites to markup their web pages and email messages).

In literature there exist many definitions of the Knowledge Graph. For example, Wang et al. [37] generalize the Graph to be a collection of relational facts represented in the form of a triplet; Ji et al. [16] state simply that KGs are directed graphs composed of entities as nodes and relations as edges; in Jia et al. [17] KG is an actual graph with entities of different types being nodes and various relations among them being edges. Paulheim [30] attributes the following characteristics to a KG: it contains real world entities and their interrelations in a form of a graph, allows for potentially interrelating arbitrary entities with each other and can cover various topical domains. Popping [33] defines the knowledge graph as a kind of semantic network that uses only a few types of relations, but also claims that additional knowledge may be added to the graph.

It is interesting to observe the trends of research on Knowledge Graph or Knowledge Schema construction in today's scientific society, especially in the app stores analysis domain. The authors have analyzed two big scholarly literature search engines: Google Scholar and Web of Science, searching for papers only in English language, for the period of the last 5 years. Analysis was divided into four steps. In the 1st step queries with particular keywords were set in each of the engines. In the 2nd step only paper titles were read, and the authors rejected those not fitting the topic completely (which did not contain either of the keywords or, even if did contain, belonged to a totally different knowledge domain). In the 3rd step the abstracts (together with keywords) of the selected papers were read, and the authors rejected the papers which were not dedicated to KGs and/or mobile applications mining and/or clustering. The 4th step consisted in reading the full papers and analyzing whether they are strictly about using a KG for analyzing and/or clustering mobile applications, whether the methodology described fits the domain selected by the authors.

In the analysis of Web of Science, the authors used the keywords in three groups: 1) TI = (knowledge graph AND app); 2) TI = (knowledge graph AND app store); 3) TS = (knowledge graph AND mobile app*). In accordance with the Web of Science search rules, TI means searching through titles, TS is a search through topics, AND is an operator meaning both keywords have to be present together, "app*" means that the system will include derivatives like "apps" and "application" into the search. Query for the first group resulted in obtaining 1 paper only; for the second group no papers were found; the third group query gave 58 papers. The paper from the first group was accepted by the authors at stages 2 (title), 3 (abstract) and 4 (content). It is dedicated to development of a knowledge-graph-based process of clustering mobile applications (see Table 1). The number of papers from the third search group was reduced to 8 in the 2$^{nd}$ step and to 1 in the 3$^{rd}$ step. After reading the paper fully in the step 4, the authors rejected it as not relevant to the selected topic.

---

[4] Google blog [https://www.blog.google/products/search/introducing-knowledge-graph-things-not]
[5] https://wiki.dbpedia.org/
[6] http://basekb.com
[7] A collaborative community maintaining structured data on the Internet [https://schema.org]

When creating a query in Google Scholar, two groups of keywords were used: 1) „knowledge graph" and „app"; 2) „knowledge graph" and „app store". Google Scholar searched for keywords both in paper titles and in paper topics. The authors searched for papers only in English language, for the period from 2017 till 2019. In the first group 509 results were shown (as for March 2019), while in the second – only 34. At the 1$^{st}$ stage the number of papers in the first group was reduced to 70, while in the second group – to 11. After the 2$^{nd}$ stage, the first group contained just 9 papers, and the second – only 1 paper. The 3$^{rd}$ step has brought the following results: group one in the end contained only 7 papers, group 2 – no papers. The papers (see Table 1) in group one are dedicated to construction of various types of Knowledge Graphs for analysis of applications content, users preferences, as well as for building clusters of applications based on their metadata.

Table 1. Knowledge graph usage in app store mining: recent literature review

| No. | Research paper | Research results |
|---|---|---|
| 1 | [3] | An algorithm that assimilates knowledge from written digital content – via modelling knowledge from digital documents into a KG. |
| 2 | [4] | A tool for conducting topic clustering of mobile applications in the search result – based on a KG that groups topic labels into topic clusters. |
| 3 | [20] | An extension to Wikidata query service, based on semantic technology, in order to make the query mechanism more reliable, flexible and useful for advanced data analysis. |
| 4 | [25] | KG usage for representing Android apps and their relations, based on the processed data retrieved from different sources (Google Play and others). |
| 5 | [26] | Automatic completion of a KG (inferring missing entities and relation types) and its application for user actions prediction. |
| 6 | [31] | Investigation of knowledge transfer into a KG, its completion and performance – in the sphere of user support in online social networks. |
| 7 | [32] | Semantic model based on a KG, applied for gathering data from heterogeneous sources into one information-based data platform. |
| 8 | [38] | Mobile application based on semantic web technology and a KG, which simplifies data processing in museum sphere. |

The authors admit that with engagement of more scholarly databases the literature search might reveal more research works dedicated to the selected topic. Yet, it can be stated that Google Scholar and, particularly, Web of Science, are widely-used search engines and contain a lot of up-to-date scientific literature. Therefore, the outcome of analysis of the selected scholarly papers allows to claim that Knowledge Graph development (and Knowledge Schema construction) for the purpose of mobile applications analysis and grouping (clustering) is not yet a very widespread research issue. This fact gives the authors space for their own suggestions and experiments.

The review of existing literature has allowed the authors to formulate the following Research Questions (RQs) as for using the data from an app store to develop a Knowledge Graph:

RQ1a. Can the data downloaded from Google Play store be represented as triples in a Knowledge Schema?

RQ1b. Are the data downloaded from Google Play store sufficient to be modeled as a Knowledge Schema?

RQ2. Are Knowledge Graph algorithms suitable for modeling the data retrieved from Google Play?

## 3. Research Method

To collect data for analysis and further usage in a KG development, Google Play store (in its website format) was selected by the authors. Google Play is one of the world's largest (and most frequently used) repositories of mobile applications. The number of apps the store currently offers is 3.6 million apps (as for May 2019[8]); there are more than 2 billion[9] monthly active devices running on Android (for which Google Play operates). Analysis of Google Play turned out to be more convenient for authors also for the reason that, except for having an application for smartphones and tablets, the store offers a user-friendly website, where all the information on apps is available. Apple App Store also has its website, but in the data retrieval process, after a few hundreds of downloaded apps, it blocks the retrieving software, preventing from massive data gathering.

In order to gather data from the Google Play pages, the authors used an automated software for websites crawling – Screaming Frog SEO Spider[10], which retrieved data, divided into 11 groups based on their type. The Screaming Frog software is one of the widely-used website crawlers. The reason why is was chosen by the authors is, in the first place, the combination of retrieval techniques it applies. First technique, based on regular expressions, used them to match 5 elements (number of downloads, last date of update, content rating, range of pricing and software requirements). The second technique used CSSPath to match the next 5 elements (developer, number of reviews, category, name, average rating). The URL was also retrieved for each application during the crawling process.

These two different techniques were necessary for the authors to get data from Google Play, because of the difference in how data are located in the website. Some of the elements are written down firmly into a website structure and are always placed in the same context (visually – in the same place in a webpage). For this type of data, the CSSPath was used. The rest of elements can change their position in website structure, in particular due to the incomplete data provided by the developer: some of the apps in Google Play do not give all the information that usually should be displayed in the store. Regular expressions technique helped to collect information published in different parts of the website.

The initial step of Knowledge Schema development is defining the knowledge domain in which it will be constructed. For this particular research the domain includes software applications terminology, particularly those for mobile devices. With reference to this domain and in accordance with the rules of KS vocabulary development, the authors have applied the RDF encoding terms from Schema.org and have added a few elements necessary for construction of a proper KS in the domain. Therefore, the entities in the Schema have the following categories: Class – can be applied for a subject or an object, defines the type of entity, especially the one that is not included in the vocabulary (e.g. User belongs to the class Person); Attribute – the predicate of each triple, indicates relationships between entity classes or between a class and a data type (e.g. Person has *name*); Type – applied for a subject, defines the type of data for a particular class (e.g. Person has *name* of data type Text). A specific rule is applied for writing down attributes in each triple: if an attribute is complex (more than one word), it is written with no spaces, each first letter capitalized (e.g. *subClassOf*, *legalName*).

---

[8] https://www.businessofapps.com/guide/app-stores-list
[9] https://www.theverge.com/2017/5/17/15654454/android-reaches-2-billion-monthly-active-users
[10] https://www.screamingfrog.co.uk/seo-spider

The Knowledge Schema based on Google Play store data is presented in Figure 1.

In order to proceed with the further steps of Knowledge Graph development, the correctness of a KS needs to be verified. Firstly, a simple reading through its visual form would be reasonable – in order to make sure all the entities are connected correctly. Afterwards, the KS can be transformed into the language of representation, which will be further read by a selected information system. One of such languages can be the Terse RDF Triple Language (called Turtle), which is a syntax for RDF. A Turtle document allows writing down an RDF knowledge graph in a compact textual form. It consists of a sequence of instructions, statements that generate triples, blank lines and (if necessary) comments. This language also guarantees compatibility with the systems that will execute queries in the Knowledge Graph. When a Turtle document is ready, it is tested. There is a need to make sure that the Classes, Attributes and Types work together properly; that a simple query would give correct results. One of the rather simple options to verify the way a KS (and its Turtle document) is constructed is using a free and open-source semantic application OntoWiki[11].

## 4. Data and Results

### 4.1. Data

In the process of app store mining, the authors have downloaded from Google Play information about 50 000 apps, which belong to 48 different categories. There is the main category Games, which includes 17 subcategories (action, adventure, arcade, board, card, casino, casual, educational, music, puzzle, racing, role playing, simulation, sports, strategy, trivia, word). In the downloaded sample games take 16,66% of total number of applications, and other apps belong to 31 different categories. Figure 1 represents shares of all categories in this sample.

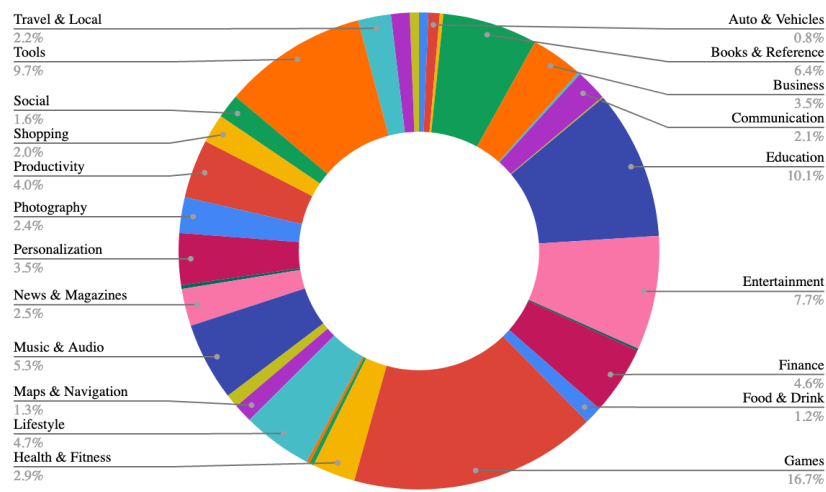

**Figure 1.** App categories in Google Play store
Source: Own app store mining

When exploring pages of the Google Play website (in order to select the parts of information for retrieval), the authors have faced the unexpected settings of data presentation (described below), that did not prevent the authors from continuing the process, yet turned into important factors that need to be taken into consideration in order to have all the data retrieved correctly.

The data in the Google Play store adapts to language settings of user's browser and operating system. During initial screening the authors have seen different settings coming from different languages. The first factor is that prices of an app itself or in-app purchases are displayed in the currency set in the web browser. The second factor: the types of content rating are different for different localization settings. The authors checked three options of language & country settings: Poland, Ukraine and US. For Poland and Ukraine, the rating is displayed in Pan European Game Information (PEGI[12]) standard, while for US the rating format comes from the Entertainment Software Rating Board (ESRB[13]) [6]. Finally, the settings for US were adopted by the authors to proceed with the data mining. Such choice can be explained by the fact that US settings are the most common to be applied by users (not only those from the USA). The third factor revealed, is that except for recommending applications on the basis of language and country, Google Play provides different search results for authorized and non-authorized users. If not logged in into Google account, a user is suggested a set of applications that differs from the one an authorized user will get (although the query in both situations is exactly the same.

---

[11] http://ontowiki.net

### 4.2. Results

For construction of the Knowledge Schema for Google Play store (Figure 1) the authors used 11 groups of data collected from the store (as described above). In accordance with the Schema.org vocabulary, all mobile applications from Google Play belong to the class Mobile Application, which, in turn, is a subclass of (attribute *subClassOf*) Creative Work – one of the generic classes in vocabulary, which covers all possible creative works (e.g. books, movies, paintings, software products, etc.). The name of an application is represented by the attribute *name* (for the class Mobile Application respectively).

Developers of applications in the Schema are represented by two classes – Person and Organization, since applications are developed not only by companies, but also by individuals. Both Person and Organization have their names (attributes *name* and *legalName* respectively) of the data type Text, and web addresses (attribute *url*) of data type URL.

Category of an application (e.g. Game, Family, Business) is described by the attribute *applicationCategory* (data type Text), while its subcategory (e.g. Adventure Game, Creativity) is the attribute *applicationSubCategory* (data type Text as well).

Class Aggregate Rating is used for two values (attribute *aggregateRating* which belongs to the class Mobile Application): the number of reviews given to an application by users (attribute *reviewCount* which has data type Integer), and the average rating of an application at Google Play (attribute *ratingValue* which has data type Number). Application content rating (age limits for using a particular application, e.g. 3+, 12+, 18+, everyone) is described by the attribute *contentRating* with the data type Text.

To give information on software requirements of an application (e.g. needs Android 4.1 and up) the authors chose the attribute *operatingSystem* with the data type Text. Additional purchases in an application (e.g. filters for photos, weapon units in games) all belong to the class Offer (attribute *offers* for class Mobile Application).

To consider the number of downloads of an application, the authors have added the attribute, which does not originally exist in Schema.org vocabulary: *downloadCount* with data type Integer. Last update (day, month and year) of an application goes with the attribute *dateModified*, which has data type Date.

Finally, for the web address of an application, the attribute *installUrl* is used (data type URL). In accordance with the Schema.org description[14] this attribute is used for a URL at which the application may be installed. In case of the URLs collected from Google Play store, these are links to pages with all the information (discussed above) about an application and the button "Install", which allows to install this application on a particular device. Thus, the authors consider the attribute *installUrl* to be suitable for this kind of web address.

---

[12] https://pegi.info
[13] http://www.esrb.org
[14] Attribute "installUrl" [https://schema.org/installUrl]

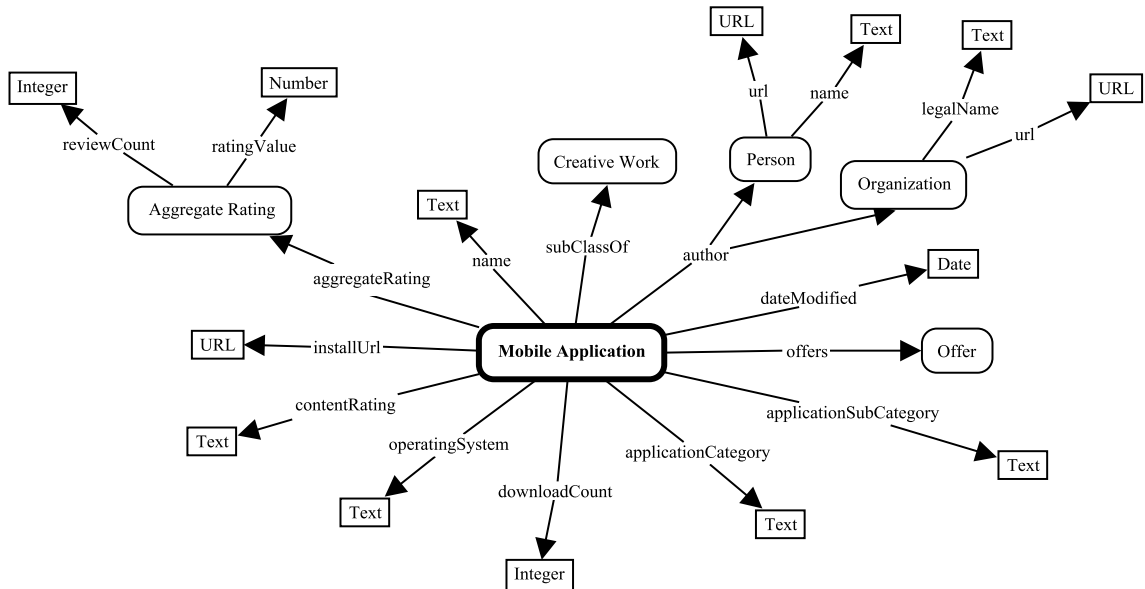

Figure 2. Knowledge Schema: Mobile Application at Google Play store
Source: Own elaboration (based on Schema.org vocabulary)

The Knowledge Schema, as stated above, is the meta-layer of a potential Knowledge Graph, and its graphical representation. It shows how the entities from Google Play are interconnected and, if constructed correctly, should be comprehensive and understandable. Based on this KS, a Knowledge Graph is supposed to conduct users queries and give information in a convenient and clear form. For instance, if a user would like to find the most downloaded app, the Graph will present search results with the apps sorted by the attribute *downloadCount*. At the same time, it may additionally inform that a particular application has the rating 4,5 and suggest seeing all the other apps with this rating value (*aggregateRating*). The rating results may also be grouped by *applicationCategory* or *applicationSubCategory*. Grouping and sorting of applications will be performed based on the instructions, written in the RDF document.

## 5. Contribution and Discussion

### 5.1. Discussion

The method, selected by the authors for Google Play mining, has allowed to retrieve a large massive of data about mobile applications. The current stage of the authors research has brought the following results. Research Questions 1a and 1b were answered. The 11 categories of data, retrieved from Google Play, were successfully transformed into classes and attributes, which form triples in the constructed KS. Moreover, these categories almost fully fit the Schema.org RDF vocabulary, selected for the KS construction. Only 1 attribute (*downloadCount*) was added by the authors. Research Question 2 was answered partially, since at this particular stage the authors do not develop the Knowledge Graph as it is but prepare its meta-layer (a Knowledge Schema). However, since the retrieved data were divided into unambiguous categories and form a KS which is clear and comprehensive, it can be stated that these data form a suitable and sufficient base for developing the Knowledge Graph itself in the next step.

### 5.2. Contribution

In recent years the usage of mobile gadgets like smartphones and tablets has grown significantly. Tablets are even being used at schools to enhance the learning process and make it more interactive and gamified [12]. It is not doubtful that work with these gadgets is connected with usage of dozens of mobile applications, which today offer a great variety of functions not only for entertainment, but also for business and educational purposes. Striving to attract users and keep them interested and satisfied, large mobile applications repositories (like Google Play, Apple App Store or BlackBerry World) offer more and more apps on their platforms. Sometimes these apps do not differ in functions and only have different interfaces or color palettes; some of them may offer additional functions do be bought in the app. Despite the perks of having a wide choice, users of app stores may face inconveniences when having to choose one application from a long list of similar ones.

The findings of this paper are as follows. Firstly, it was revealed that today's literature on Knowledge Graph usage does not yet contain much research on developing a KG for the needs of mobile applications analysis and clustering. It means there is still a field for development of algorithms that could improve search results in app stores, particularly with usage of a KG as tool for apps clustering. Secondly, the authors have proved that a website crawling software tool is capable of retrieving from an app store a large sample of data, which by content and format will be suitable to be used for KG development. Thirdly, the authors constructed the Knowledge Schema that models the mobile application concept and shows interrelations of data pieces that describe any mobile app. The Schema is built on the basis of apps from Google Play store, which is currently one of the world's leading online store for mobile applications. Knowledge Schema is the meta-layer of the Knowledge Graph. Schema construction is a significant step towards KG development. Graph is a collection of entities, which are processed in order to run the query of a user and present the necessary information. These entities need to be properly connected in the Schema first, to make it work.

### 5.3. Future Research

The authors have divided the process of Knowledge Graph development into a few stages. Stage 1 (presented in this paper) consists in retrieving the necessary data from an app store and forming a dataset that will be then used in a KG. It also includes definition of the knowledge domain and RDF vocabulary belonging to it; and construction of a Knowledge Schema in the end. Once the Schema is built, stage 2 can be started.

At stage 2 the authors are going to examine the Google Play dataset with the help of sentiment analysis tools. Analysis of the sentiment of users' reviews will result in obtaining a new category that characterizes Google Play applications – tonality of users' reviews. Thus, an application would be characterized by its rating, number of reviews and the emotions expressed in these reviews (e.g. like, dislike, admire, hate, etc.).

Stage 3 presupposes processing of a draft Turtle document, which reflects the structure of the future Knowledge Graph. In accordance with the rules, such document contains triples (from the KS), links to particular elements of the Google Play dataset and specific instructions on data processing. At this stage multiple verification tests are conducted to eliminate mistakes and make the KG perform correct queries. The OntoWiki open-source application is one of the tools the authors suggest using for verification.

Eventually, at stage 4 the authors are going to obtain a functional Knowledge Graph. Such Graph is supposed to serve as a kind of a recommender system for users of the Google Play store: a system that will not only divide applications into categories, but also provide users with information about apps' quality (judging by other users' reviews) and suggest (recommend) other applications (with the similar rating, of the similar category, with the similar reviews tonality, etc.). It is of no doubt that implementation of the KG into a commercial app store like Google Play (or any other) is not a simple task. A real thing for the authors to do would be development of a separate platform operating on the basis of the Knowledge Graph and serving as a system that provide users with recommendations on mobile application at the app store. Once such a platform is constructed, it is reasonable to test it on a small sample of users – not only to check whether it operates correctly, but also to reveal how useful it turns out to be for someone in search of a mobile application.